# The first alumni donation in 1880 in Japan: social image and the open-academic record system

Eiji Yamamura*

Department of Economics, Seinan Gakuin University, Fukuoka, Japan

*Corresponding author's email: yamaei@seinan-gu.ac.jp

ABSTRACT

In 1880, Keio, a private school in Japan, was in jeopardy of being closed. To cope with the situation, the school first created a fundraising campaign during the 1880–90 period. The school was established in 1857, and since 1861, the list covering all students' academic record has been distributed not only to teachers but also to all students. Individual-level historical academic record was integrated with the list of contributors. Using the data, we compared persons who had learned in Keio before and after the system was introduced. The main findings are presented as follows: first, graduates who share the academic record are more likely to contribute, and their amount of donation is larger; second, the class size is negatively correlated with the likelihood of contribution and with its amount; and third, academic performance, as shown in the list, is positively correlated with the likelihood of contribution but not with the amount of donation, using a sub-sample of those who shared the list. The introduction of the system strengthened the community network and role of social image shared by the members. This resulted in a successful fundraising for the school, an unprecedented feat in the history of Japan.

2

# 1. INTRODUCTION

The oldest private university in Japan, Keio Gijuku, was established in the mid-nineteenth century. Within several decades, many gentlemen who had learned at Keio had become business elite because Yukichi Fukuzawa, the founder of Keio, aimed to foster the competent human resource to form a group of business leaders in Japan (Tamaki, 2001, 2002). However, exogenous shocks, such as the civil war in the nineteenth century, drastically reduced the student population. Without support from the Japanese government, Keio was in the predicament of being closed at the end of 1870s. To pull through the difficult situation, Fukuzawa mounted a fundraising campaign to establish the Keio Gijuku Foundation and to run the school independently.

Prior to the jeopardy, from 1861, the list including all students' academic record has been distributed not only to teachers but also to all students in Keio. The open-academic record system is considered to reduce information asymmetry among members of Keio Gijuku and to share the function of community mechanism among the members of Keio Gijuku (Greif, 1994, 2006, 2012; Greif & Tabellini, 2017; Hayami & Godo, 2005). We examine how the introduction of the system is a prerequisite to succeed in the first school fundraising campaign in the history of Japan.

The Keio alumni network formed by the business elite fulfils a vital role vis-à-vis gathering large amount of donations sufficient to maintain Keio Gijuku. This is the first successful school fundraising in Japan (Tomura, 2011). However, here, a question arises. Why were they motivated to contribute to the donation? Several previous studies have attempted to empirically investigate this situation. University reputation and quality give alumni an incentive to contribute (Baade & Sundberg, 1996b; Cunningham & Cochi-Ficano, 2002; Meer & Rosen, 2009b). The probability of donation depends on sports-related variables such as alumni sports experience in university and university's success with regard to sports (Baade & Sundberg, 1996a; Holmes et al., 2008; Tucker, 2004; Turner et al., 2001; Wunnava & Lauze, 2001). The effect of sports experience is only observed for male alumni but not for female ones (Meer & Rosen, 2009b). Satisfaction level during their learning at university is positively correlated with alumni giving (Clotfelter, 2001, 2003). Alumni contributed because they expected that donation would raise the probability of their child's admission (Meer & Rosen, 2009a). The effects of financial aid received as students varied with the type of aid (Marr et al., 2005; Meer & Rosen, 2012). However, existing studies deal with the alumni giving in the present day when the system of alumni donation has been firmly established in the developed Western countries.

In the nineteenth century, researchers used the quantitative data of school attendance in England and Scotland to examine the effect of government intervention on private schools (Anderson, 1983; West, 1970, 1971). The school attendance rate for males was 80 per cent in 1871 in Scotland (Anderson, 1983), whereas the rate was only 39.9 per cent for males in 1971 in Japan (Ministry of Education in JAPAN, 1981).[1] The large gap of the average educational levels existed between Japan and the United

---

[1] The data of Japan were open to public in the Official website of 'Ministry of Education, Cultures, Sports Science and Technology', JAPAN.

States in the early Meiji period, although the gap was significantly closed by 1990(Godo & Hayami, 2002). Furthermore, even in the mid-nineteenth century, the higher engineering school provided the commercial education to supply qualified labour to a business world in European countries (Passant, 2019), whereas Japan's education system had not subdivided in 1880. Arguably, the education system had been less matured in Japan than in the Western countries when Keio Gijuku started the fundraising campaign in 1880. At least, only young boys brought up in wealthy families could enter Keio Gijuku, and then, the alumni were naturally likely to be the business elite, as Fukuzawa expected (Tamaki, 2002).

As argued in comparative institutional analysis, cooperation is sustained by clan and corporation in China and Europe. Further, organisation is managed to follow loyalty to kin in China and to generalised morality in Europe (Greif & Tabellini, 2017). Asian society is characterised by the vertical interpersonal relationship within a tightly knitted community(Hayami & Godo, 2005). The incentive mechanism depends on institution that varies with to not only location but also era (Greif, 2006). Motivation in connection to alumni donation in the nineteenth century in Japan might be different from what obtains in Western countries today. Intrinsic motivation formed the self-image to be the key determinant of an individual's behaviour (Benabou & Tirole, 2003; Bénabou & Tirole, 2006). As an example, prosocial motivation leads individuals to contribute to donation. More recently,

---

https://www.mext.go.jp/b_menu/hakusho/html/others/detail/1317590.htm (Accessed on Sep 26, 2023)

economists have paid great attention to 'social image'(Bursztyn & Jensen, 2017). The social image shared by a group gives its members an incentive to maintain their behaviours to meet the image, for instance, when people vote in an election (Dellavigna et al., 2017; Funk, 2010; Gerber et al., 2008). This provides clues to analyse the reason that alumni contribute to fundraising. During school life, students may be motivated to demonstrate their stylishness to colleagues. However, stylishness changed according to the 'social image' shared by the group whereto a student belonged. Students would not endeavour to learn if their colleagues consider it 'stylish'. Therefore, forming the social image is considered as the commitment device. The open-academic record system fulfilled a critical role in forming the 'social image' which was strengthened by Keio inhouse media.

We constructed the historical individual level data by connecting three different individual level datasets, school enrolment list from 1848 to 1879, academic record in the list from 1871 to 1879, and list of the contributors to the fundraising from 1880 to 1890. We used the data to compare those who learned in Keio before and after the system was introduced. The major findings include, first, graduates who share academic records are more likely to contribute and their amount of donation is larger; second, the class size is negatively correlated with the likelihood of contribution and its amount; and third, academic performance shown in the list is positively correlated with the likelihood of contribution but not with the amount of donation using a sub-sample of those who share the list. The introduction of the system strengthened the community network and role of social image shared by the members.

## 2. OVERVIEW OF KEIO GIJUKU AND EDUCATION IN JAPAN

At the end of Edo era, a Japan private school, Keio Gijuku, was established in 1858[2]. Subsequently, various private schools were established in the 1860s and 1870s, but they closed down. Conversely, Keio Gijuku has successfully survived and extended to integrate primary school into university. In present-day Japan, Keio Gijuku has become the oldest private university. However, similar to other now-defunct private schools, Keio Gijuku was in a predicament from end of the 1870s mainly due to the reduction of students as a consequence of the civil war (Seinan War) that occurred in 1877 (Keio Gijuku, 1958; Motoyama, 1994). The Japanese government has not supported private schools, rather favouring public schools[3].

To cope with the jeopardy, Fukuzawa started a fundraising campaign to create the 'Keio Gijuku Foundation' in 1880, becoming the first private school fundraising in the history of Japan (Keio Gijuku, 1958). Furthermore, the foundation aimed to not only overcome the financial difficulty but also maintain a cash reserve for investing for the future (Nishikawa, 1999). This caused Keio to survive and be independent of the government. Consequently, Keio was considered as the most successful school in nineteenth century Japan in fundraising (Tomura, 2011). There are critical factors to succeed to collect the donation. From 1871, Keio has made the list of all students' academic record covering from the primary school ('Yochisha') to the highest grades. Furthermore, the list called 'Kindahyo' was distributed to all members of Keio—teachers and students. The list has published every academic term, and so three times in a year because there were three academic period. In the list, students appeared in order of merit in each class[4], in addition to the number of days of attendance, score of achievement test about various subjects such as English, mathematics, law, economics, and so on[5]. Moreover, the number of students in a class were usually 10–20 in the studied period before 1880. Naturally, during school life, students know about each other with respect to not only characteristics but also detailed academic record. Among the students, an individual's information was 'symmetric' from the viewpoint of economics.

---

[2] There were no female students in Keio in the Meiji era, and the first female students emerged in 1948 just after World War II . At least in the Meiji era, the gender-mixed school had not been established in Japan. The official website of Keio Gijuku University. https://www.keio.ac.jp/ja/contents/stained_glass/2013/278.html. Accessed on Aug 26, 2023.

[3] No subsidy has been given to Keio Gijuku even though its founder, Yukichi Fukuzawa, frequently submits petitions. Further, a conscription ordinance was revised in 1883, causing private schools to lose the privilege of exemption from conscription (Motoyama, 1994).

Furthermore, Fukuzawa raised the slogan 'Ji-sha Kyoryoku' for members of Keio in the written rule of the Keio Gijuku Clan, 'Keio Gijuku Sha-chu no Yakusoku' in 1871 (Keio Gijuku Fukuzawa Research Center, 1989a).[6] 'Sha' in 'Sha-chu' and 'Jisha' means clan. In the slogan and the rule, Fukuzawa emphasised 'cooperation within clan of Keio members' to tighten the solidarity of Keio

---

[4] The list was made from 1871 to 1898. During the period, the number of classes changed. However, in the 1871–80 period, which was prior to the fundraising, the number of classes was 13, although the class size (students in a class) became smaller after 1877.
[5] Subjects differed according to grades, and the curriculum changed frequently.
[6] Fukuzawa advocated 'self-reliance and self-esteem' for the rest of society.

members comprising students and alumni.

This shares similarity with the organisation of China based on clan (Greif & Tabellini, 2017). This generated 'social image' was shared among Keio members. Accordingly, they should unite strongly against the predicament with which Keio school encountered. Especially, Fukuzawa urged the necessity of 'Jisha Kyoryoku' directly before starting the fundraising campaign in 1880.[7] In addition, Keio has published an in-house newsletter and announced the commencement of the fundraising campaign to cope with the jeopardy. Names of alumni who donated were reported consecutively therein. The social image of 'clan cooperation' promoted by the media was considered to give the alumni a strong incentive to contribute their donations.

Fukuzawa formed the network with lower grade private schools all over Japan. Students could enter Keio if they obtained the recommendation from the school they had learned because there was no entrance examination in Keio before 1890. Especially, Fukuzawa sent teachers to his hometown's school (Nakatsu school) to build strong connections with the hometown's education to optimise human resources and facilitate confidence-building in the future. Fukuzawa endeavoured to cultivate reliable interpersonal relationships to not only gather students but also supply Keio graduates to private leading firms. Alumni networks were effective to match supply and demand in Japan's labour market (Chiavacci, 2005). Thanks to Fukuzawa's contribution, especially the Keio alumni network, Mita-kai,

---

[7] Source: 'Encyclopedia of Keio Gijuku' in the official website of Keio Gijuku University. https://www.keio.ac.jp/ja/about/history/encyclopedia/36.html. (accessed on Sep 27, 2023).

has been dominant in the business world until present days[8].

Fukuzawa went on official trips abroad three times as a member of the mission to the West and was influenced by the Europe and United States. He planned to create the first business elite from Keio Gijuku (Tamaki, 2002). Because of his educational philosophy, Keio focused on practical studies exported from the modern western countries and fostered capable human resource to supply labour for dominant firms, such as Mitsui and Mitsubishi Zaibatsu (company syndicates) (Tamaki, 2001). He was a successful entrepreneur and advisor to Mitsubishi and Mitsui Zaibatsu (company syndicates). In the late nineteenth century, some leading Japanese firms adopted the modernised and scientific management skills. Alumni gathered the personal information from his Keio Gujuku and then recruited competent students. As an example, Ryohei Toyokawa was the cousin of Yataro Iwasaki, who was the founder of the Mitsubishi Zaibatsu. He learned at Keio and then entered Mitsubishi. Through the Keio alumni network, he recruited the highly reputable Heigoro Shoda to Mitsubishi (Kobayashi, 2007). Shoda learned commercial bookkeeping in Keio and then applied it to the management of firms under Mitsubishi. Owing to the scarcity of highly educated human resources in the Meiji era, Keio could supply very valuable labour to private firms. Especially, Zaibatsu firms demanded the reliable and promising Keio students through 'clan connection'. Closing Keio would reduce labour supply, thereby hampering firm growth. The executives of Zaibatsu firms have strong motivation to donate. Therefore,

---

[8] 'Mita' is the name of address where the main campus of Keio Gijuku is located.

donations could be collected from Keio alumni who were very wealthy business elites.

## 3. HYPOTHESES

The open-academic record system enabled Keio students to share characteristics of all members, thus reducing information asymmetry between them. Naturally, the relations between students became more intimate, hence forming the Keio clan. Even after graduation, their tight bond persisted to enjoy the benefit from the Keio clan through the alumni network. As demonstrated in Fig.1, the donations were accepted for about 10 years, from 1880 to 1890. A Keio alumni magazine made a major fundraising campaign. The magazine published the names of donors and the amount of money they had raised from 1880 until the deadline in 1890. The alumni knew that their fundraising efforts would be publicized in the Keio community. 'Social image' to reciprocate and contribute to Keio might be shared if they pursue benefit from the long-term relationship. Accordingly, we propose *Hypothesis 1*,

*Hypothesis 1: Keio alumni who have learned under the open-academic record system are more inclined to contribute to the fundraising.*

High ordinal academic ranks improved the cognitive skills measured by academic performance (Denning et al., 2023; Elsner et al., 2021; Elsner & Isphording, 2017; Murphy & Weinhardt, 2020). Further, the rank improved non-cognitive skills such as conscientiousness (Pagani et al., 2021) and self-confidence (Isozumi et al., 2021). However, cooperativeness is more important to improve teamwork and firm efficiency from the viewpoint of economics. Moreover, cooperativeness is considered to give an incentive to contribute to fundraising, especially if one belonged to tightly knitted interpersonal relations. To examine whether the ordinal rank improves the cooperativeness, we raise *Hypothesis 2*:

*Hypothesis 2: The higher the ordinal rank in the school days, the more alumni are likely to contribute to the fundraising.*

As explained in the previous section, Zaibatsu firms demanded highly educated Keio students who learned the management science to improve efficiency and firm performance. The executives of Zaibatsu firms would lose their source of labour supply and expected profit if Keio is shut down. To pursue long-term profits and firm growth, executives in Zaibatsu firms have an incentive to donate. This leads us to propose *Hypothesis 3*.

*Hypothesis 3: Workers in Zaibatsu firms are more likely to donate.*

## 4. DATA AND METHODS

### 4.1. COLLECTING AND CONSTRUCTING DATASETS

We use the three different individual-level datasets that are included in series of Fukuzawa-related historical documents 'Fukuzawa Kankei Bunsho' reserved as micro-film[9]. First, list of freshmen-students, 'Nyu-sha cho', contained the name and social status 'Samurai or Ordinary citizen', certifier's social status, position in the family 'the head of a family or not', and address of hometown (Keio Gijuku Fukuzawa Research Center, 1989b). Second, the academic record list, 'Gakugyo~Kinda-Hyo', included the academic record of all Keio students from 1871 to 1898 (Keio Gijuku Fukuzawa Research Center, 1989c). Ordinal rank in the class, achievement test scores for various subjects, and attendance days are included. However, we often found blanks, especially with regard to the certifier's social status. Third, the list of contributors to the fundraising in 1880 includes the contributor's name and amount donated (Keio Gijuku Fukuzawa Research Center, 1989d). The fundraising started in 1880 and the contributor applied to donate and decided an amount of donation to be contributed within several years. Ten years after, the list has been published since 1890, showing the amount of donation that one has contributed (see Figure 1).

---

[9] Besides the datasets used in this study, the series of documents included various kinds of historical documents and data from the establishment of Keio to 1901, when Fukuzawa passed away.

< Insert Figure 1 >

Identifying individuals is difficult when we connect three datasets because individuals' names often change for a certain number of students. There seemed to be various reasons to change name. Some students entered a family as an adopted son and thus changed their names. Besides, the revision of a conscription ordinance deprived the privilege of exemption from conscription from students of private school (Motoyama, 1994). To evade conscription, some students changed their name to not be identified. Further, the way of changing name varied. Even if an identical individual is listed in the three datasets, the family name is different, first name is different, and both family and first names are different between datasets. Therefore, using various sources and documents, individuals with different names are identified.

## 4.2. COMPOSITION OF THE DATASETS

<Insert Table 1>

Table 1 presents the composition of the datasets. We assume that students could not apply to fundraising, we limited the sample of alumni who learned at Keio after its establishment in 1858 to the year of the fundraising campaign in 1879. Actually, nobody who was a student in 1880 contributed, and so the assumption is reasonable. Further, before 1865, there was no notion of graduation and so nobody graduated from Keio. So, the total number of those who learned at Keio until 1880 were 3,503 that were registered in the list of enrolled students. This is called Sample (A) in this study. As explained in the next section, Sample (A) was used for the estimation to assess those who contributed.

Sample (A) comprised two groups: the group of those who learned before the open-academic record system during the 1858–70 period, while the other group encompassed those who learned after the introduction of the system during the 1871–89 period. The former and latter observations are 720 and 2,333, respectively. According to the rule, all students automatically appeared in the list if they learned under the system. However, actually, 23 per cent of students (547) did not appear in the academic list. So, the academic record was available for 1,786 students. Individuals' academic performance is available three times every year during the period the students learned. As explained later, the average ordinal rank of all terms they learned was calculated; thus, the number of observations were equivalent to number of individuals. In the list of enrolled students, for some students, basic information such as age, birth year, and entrance year were not filled.

Among the alumni in Sample (A), 100 individuals donated, representing only 3.3 per cent. Sample (B) was used to estimate the donation amount. Similar to Sample (A), Sample (B) can be divided into two groups to learn before and after the system. Some of the items of information are unavailable.

4.3. MEASUREMENTS

< Insert Table 2 >

Table 2 describes the definition of variables used in the estimation, and its basic statistics. The key variables are as follows: to measure donation behaviour, we use the donation dummy that is 1 if the alumni have contributed, and otherwise 0. In the academic record list, scores of the achievement test of various subjects and number of attendance days are used. However, the set of subjects not only varied according to grades but also changed frequently to time points even for the same grade. Furthermore, days of attendance depended on the length of term that also varied. However, ordinal rank in the class whereto students belonged reflected the combined score of all subjects and days of attendance. In addition, many recent studies show that the ordinal rank is valuable and useful as measure of academic performance (e.g., Denning et al., 2023; Elsner et al., 2021; Elsner & Isphording, 2017; Isozumi et al., 2021; Murphy & Weinhardt, 2020; Pagani et al., 2021). Hence, the ordinal rank in the class is used as a measure of academic performance in this study. However, academic rank in a class is influenced by class size. The first place among 100 students means top 1 per cent, while the first among 2 students means top 50 per cent. Therefore, we should control the number of students in the class. In the data, the mean value of '*Class size*' (number of students in class) is 25.1, and its standard deviation 11.2 in Sample (A). Thus, the variation between classes is fairly large. Therefore, the rank is divided by number of students in class. Hence, the bottom place is equal to 1 whereas the top place is the nearest to 0 in the class. Further, we calculate its mean value to determine the '*Academic Ranking*'. We should interpret it carefully. The smaller the value, the better the performance.

Apart from academic performance, the degree of loyalty to Keio is measured by several variables. The degree of attachment to the Keio school is measured by number of terms a student has learned, and so length of periods in Keio (*Period Length*). Low performance students retired earlier whereas the excellent students could skip grades to higher ones and complete the education program earlier. Thus, *Period Length* is unlikely to reflect academic performance. However, attachment may be stronger if *Period Length* is larger.

The most important variable is, *Academic Record,* that is 1 if the alumni have registered in the academic list during their school days, otherwise it is 0. Its mean values are 0.51 and 0.62, implying that 51 per cent and 62 per cent of the alumni share the open-academic list for Samples (A) and (B), respectively. The difference between the samples implies that those who have donated are more likely to share the list during their school days. Fukuzawa's hometown was Nakatsu, wherefrom many students entered Keio through strong connections therewith. *Nakatsu* is 1 if the alumni are from the same hometown as Fukuzawa, otherwise 0. Its mean values are 0.009 and 0.09 for Samples (A) and (B), respectively. Contributors were more likely to come from Fukuzawa's hometown, 10 times larger than the whole sample.

In the Western countries, the effect of upper status in the nineteenth century persisted until modern society (Clark et al., 2020; Clark & Cummins, 2015). Directly after the Edo era, the effect of social class continued to exist, although their privileges were deprived. Former samurai people belonged to the upper class in the Edo period. They were called 'Shizoku' in the Meiji period, although their privilege has been eliminated (Gordon, 2014). To measure the status, we use *Samurai,* that is 1 if one belonged to 'Shizoku', otherwise 0. Its mean values are 0.43 and 0.38 for Samples (A) and (B), respectively. The rate of upper class is 38 per cent, 5 per cent lower than that of the whole upper class alumni.

Detailed individual information such as work and job cannot be obtained for most of alumni.

However, contributors were notable persons, and so their information can be collected from various sources. Hence, classification of work *Zaibatsu Work* and *Education Work* are included in Sample (B) but not Sample (A) because detailed information cannot be obtained for most of alumni. *Age* is the individual's age in 1880, when the fundraising campaign started. Year of entering Keio is *Entrance Year*. *N.A Age* is the dummy variable to show observations of age being unavailable, while *N.A Entrance Year* is the dummy variable to show observations of the entrance year being unavailable.

## 4.4. PRELIMINARY ANALYSIS

<Insert Table 3 >

Based on Sample (A), Table 3 shows the mean difference test of the *donation dummy* to compare the probability of contributing between groups. The mean value is 0.035 for those have learned under the open-academic record system, that is larger by 0.013 points than others. The mean value is 0.29 for those from Nakatsu, that is larger by 0.264 points than others. Surprisingly, 29 per cent of Nakatsu alumni contributed, implying that a social group with common ties to the hometown was significantly motivated to obey the hometown's predecessor, Fukuzawa's, claim. These differences are statistically significant. However, there is no significant difference between *Samurai* and others.

<Insert Figure 2>

<Insert Table 4>

Based on Sample (B), Table 4 indicates the mean difference of the logarithm of the amount of donation. As demonstrated in Figure 2 (a), the amount of donation is concentrated at the bottom band 1–200 yen, although some outliers were in over 1800 yen. In Figure 2 (b), the distribution of its log-form of donations is more similar to normal distribution. So, we use log-form values in estimations. There is no difference between generations under the open academic list and others. However, we observe statistically significant differences in the following comparison. Contributors from Nakatsu provide larger amount of donation than others. Zaibatsu workers contributed larger amount of donations than others. Business elites working at Zaibatsu firms are wealthier and so can afford a larger amount. In contrast, samurai contributors' amount of donations was smaller than others, reflecting that their annual income fell by from 10 to 75 per cent after government converted their stipends into bonds in 1876 (Gordon, 2014).

<Insert Table 5>

Using Sample (A), Table 5 compares the academic related variables between contributors and others. We should pay careful attention to *Academic Ranking* because a larger value means a lower rank. The contributor's ordinal rank is higher, his class size is smaller, and his period of attendance at school is longer than others. All these differences are statistically significant. Contributors tended to exhibit better academic performance: his class size promoted collective action, and length of attendance indicated his attachment to school.

# 5. ESTIMATION METHODS

Behaviours of contributing to fundraising can be decomposed into two stages. In this setting, the Heckman-type sample selection model is widely employed. Accordingly, we also used the Heckman model. In the first stage, using Sample (A), we can assess the decision-making about whether they donate or not. Here, we can obtain estimation results effect of key variables on probability of donation. In the second stage, using Sample (B), we investigate the influence of variables on amount of donation. For convenience of explanation, we introduce the first stage and then the second stage model separately, although actually estimations in the first and second stages are conducted in a model.

## 5.1. FIRST STAGE

The estimated model is described as follows:

$$Donation\ dummy_i = \alpha_0 + \alpha_1 Academic\ Record_i + \alpha_2 Period\ Length_i + \alpha_3 Nakatsu_i \\ + \alpha_4 Ln(Class\ size)_i + \alpha_5 Academic\ Ranking_i + \alpha_6 Samurai\ Class_i + \alpha_7 Age_i \\ + \alpha_8 Entrance\ Year_i + \alpha_9 N.A.Age_i + \alpha_9 Unregistered_i + \varepsilon_i$$

In this formula, *Donation dummy i* represents the dependent variable for individual *i*. Its value is 1 or 0, and so probit estimation is conducted as the first-stage in the Heckman model. The regression parameters are denoted as α, and the error term is $\varepsilon_i$.

From *Hypothesis 1,* the predicted sign of parameter of $Academic\ Record$ is positive. Further, attachment to Keio and hometown network might increase motivation to contribute, and so we expect the parameters of $Period\ Length$ and *Nakatsu* to be positive. According to collective action theory, the sign of $Ln(Class\ size)$ is predicted to be negative (Olson, 1971).

*Hypothesis 2* leads us to anticipate the parameter of $Academic\ Ranking$ to be negative. $Samurai\ Class$, $Age$, and $Entrance\ Year$ are included as control variables. However, as shown in Table 1, *Age* and *Entrance Year* are unavailable for some individuals. To avoid selection biases, we include dummies of *N.A. Age* and *N.A. Entrance* to consider how these individuals' probability of contribution differs from individuals whose $Age$ and $Entrance\ Year$ are available. Similarly, the dummy of *Unregistered* is incorporated to control for the contribution probability of alumni who have learned at Keio in the period of the open-academic score system but whose records were unregistered in the list.

## 5.2. SECOND STAGE

The second-stage function is as below:

$$Ln(donation)_i = \beta_0 + \beta_1 Academic\ Record_i + \beta_2 Period\ Length_i + \beta_3 Nakatsu_i$$

$$+ \beta_4 Ln(Class\ size)_i + \beta_5 Academic\ Ranking_i + \beta_6 Samurai\ Class_i + \beta_7 Age_i$$

$$+ \beta_8 Entrance\ Year_i + \beta_9 N.A.Age_i + \beta_9 Unregistered_i + u_i$$

As commonly employed in the Heckman sample selection model, dependent variables, $Ln(donation)_i$, is continuous, and so ordinary least squares (OLS) estimation is conducted. The expected sign of key variables, *Academic Record* and *Academic Record* , used to test *Hypothesis 1 and 2* are equivalent to the first stage and so positive and negative, respectively. Besides, *Period Length, Nakatsu*, and *Ln(Class size)* are also predicted to take a positive sign.

From *Hypothesis 3, Zaibatsu Work* is predicted to take a positive sign. Different from the first stage, *Unregistered* is not included because relevant individuals in Sample (B) are only two. *Unregistered* enables us to avoid the collinearity problem between the first and second stages.

## 6.  RESULTS

### 6.1.  DECISION TO CONTRIBUTE

<Insert Table 6>

Table 6 reports the estimation results of the probit model. Numbers without parentheses are marginal effects. The full model result is reported in Column (1), showing results of the first stage. In addition, for robustness check, the results of alternative specifications are given in Columns (2) and

(3).

*Academic Record* shows the positive sign and statistical significance at the 1 per cent level in all columns. Marginal effect varies from 0.095 to 0.130, which implies that the probability that alumni who have learned under the open-academic record system contribute is larger than those prior to the system by 9.5–13 per cent. This strongly supports *Hypothesis 1*. Similarly, *Period Length* and *Nakatsu* indicate the expected positive sign, while being statistically significant in all columns. Both of their attachment to Keio and tight hometown network are effective to increase the probability of alumni contribution. The marginal effect of *Nakatsu* ranges from 0.128 to 0.208, which means that alumni from Fukuzawa's hometown are more likely to donate by 13–21 per cent more than others. Hence, the effect is larger than *Academic Record*. The marginal effect of *Period Length* is 0.002–0.003, suggesting that an additional term one has learned increases the probability of contribution by 0.2–0.3 per cent. Usually, there are three terms in a year. So, the probability rises nearly 1 per cent by an additional year of learning at Keio. As for *Ln (Class size),* we observed the expected negative sign and statistical significance at the 1 per cent level in all columns. This is consistent with the Olson's collective action hypothesis. Its marginal effect ranges from −0.013 to −0.021. This can be interpreted as implying that a 1 per cent increase in the number of students in a class leads to a 1.3–2.1 per cent fall in the probability of alumni contribution. The effect is sizeable.

Consistent with *Hypothesis 2, Academic Ranking* produces the negative sign and statistical significance at the 1 per cent level in Columns (1) –(3). Its marginal effect varies between −0.049 and −0.070. The value 1 means the bottom rank because the value is (rank/number of students). In our interpretation, the probability of top-place student's contribution is higher by approximately 4.9–7 per cent than the bottom-place one. The order effects is relatively modest compared with effect of *Academic Record* and *Nakatsu*. This reflects the real situation of Japan in the nineteenth century. In Japan, directly after the Meiji restoration, and thus an open country, a market mechanism where anonymous individuals exchanged was unlikely to function well. Necessarily, interpersonal effect such as social image and community mechanism exerted a larger influence on an individual's decision-making than educational achievement.

## 6.2. AMOUNT OF DONATION

<Insert Table 7>

We turn to Table 7 to consider the results of amount of donation. Numbers without parentheses are coefficients in OLS estimation in the second stage. Results of three different specifications are reported, but in the first stage, we use only the full model that is equivalent to Column (1) of Table 6. However, the estimation results of Table 7 hardly change if other specifications are employed in the first stage.

*Academic Record* shows a significant positive sign in all columns. Especially, in the full model of Column (1), marginal effects are larger than other specifications, with a statistical significance at the 1 per cent level. The marginal effect in Column (1) is 3.00, implying that alumni who have learned under the open-academic record system contribute four times a larger amount than others. The difference of donation size between groups before and after the system is distinctly and surprisingly large. The combined results of Tables 6 and 7 consistently support *Hypothesis 1*. The expected significant positive sign of *Nakatsu* only appears in Column (1), and so its effect is not robust when it comes to amount of donation. Further, no statistical significance is observed for *Period Length*. In contrast, for *Ln (Class Size)*, we observed the predicted negative sign with statistical significance in all estimations. Additionally, values of its coefficients are between −0.700 and −0.896. Both of the dependent and independent variables are in log-form, and this values indicates elasticity in terms of economics. Hence, the results indicate that a 1 per cent smaller class size increases the donation amount by 0.7–0.9 per cent. This is remarkably large. Considering the results of *Ln (Class Size)* in Tables 6 and 7 jointly leads us to argue that group size in school days fulfil a crucial role in the success of the fundraising from the perspective of collective action hypothesis in political economy.

*Academic Ranking* does not show any statistical significance, and its sign is not consistent. This possibly correlated with *Ln (Class Size)* and, hence, suffers the collinearity problem. However, no result of *Academic Ranking* shows statistical significance even if *Ln (Class Size)* is deleted from the second stage, although the results are not reported[10].

*Zaibatsu Work* consistently shows the positive sign and statistical significance in all columns. Values of coefficients range from 0.759 to 0.907, indicating that Zaibatsu business elites' amount of donation is 76–91 per cent larger than other alumni. The difference in donation amount between the Zaibatsu elite and others is remarkably large. This firmly supports *Hypothesis 3*.

# 7. DISCUSSION

The social image fulfils a vital role in determining individuals' decision-making and, consequently, their behaviours if these are open to other group members (Bursztyn & Jensen, 2017). The mechanism is relevant to the community responsibility system (CRS) functioned for merchants of Mediterranean in the late medieval period between the eleventh and fourteenth century (Greif, 1994, 2006, 2012). Under the CRS, we assume that members of 'Community A' renege on their contract towards members of another 'Community B' that imposes cost on members of 'Community A' to motivate same 'Community A' to punish the cheater and compensate for the damage. The success of the first fundraising campaign in 1880 helped build the Keio foundation and strengthened the sense of camaraderie among Keio students and graduates, and this has persisted until the present day. In the business history in Japan, Mita alumni groups took root in various industries and regions, and their members shared norms based on the spirit of 'Ji-sha Kyoryoku', initially advocated by Fukuzawa, the founder of Keio.

---

[10] The results are available upon request for the corresponding author.

Not only in fundraising but also in business transaction, a member of Mita alumni 'A' is less likely to cheat a member of 'Mita alumni B' because of 'CRS' *a la* Keio Gijuku. Accordingly, transaction cost is very low between different 'Mita alumni' groups. Thus, Keio's 'social image' maintained the reciprocal relationship between 'Mita alumni' groups, causing the Keio clan to overwhelm the alumni groups of other universities in the business world. Maintaining Keio's prosperity depends on whether the Keio clan's 'collective action' succeed. The success of 'collective action' requires the group to be small (Olson, 1971). The most prestigious group in the Keio clan is that who have learned from the Yochisha (Keio primary school) to university. Entering the 'Yochisha' is very difficult if parents of the applicants do not have connections with influential persons of the Keio clan. In other words, graduates from the Yochisha ordinarily tend to be in prestigious positions such as business leader to have strong connections with influential persons. Put differently, the connection may be inherited from family line, and so outsiders would find it difficult to enter the clan unless the applicants are distinctly excellent and talented. In addition, the older ones belonged to Keio, especially the more they exhibited loyalty to Keio. Furthermore, several 'Yochisha' graduates are far smaller than the rest of the Keio members, and so the Keio clan's 'collective action' is more able to succeed if 'Yochisha' graduates are leaders of Mita alumni groups. Naturally, the Keio alumni network continues to strive from one generation to another, and they have played a leading role in Japan's business world.

The effect of class size is interpreted from the viewpoint of collective action. However, the effect can be differently interpreted. Many studies have revealed that a smaller class size leads to improved cognitive skills (Jepsen & Rivkin, 2009; Krueger, 2003; Rivkin et al., 2005). This increases the income level in adulthood. Higher income earners and wealthy people are more inclined to contribute to fundraising (Clotfelter, 2001; Harrison, 1995). Unfortunately, we cannot identify which channel is more critical because experimental settings or valid instrumental variables are necessary but not available. However, at least, the collective action hypothesis is supported by the anecdotal evidence of Yochi-sha dominance in Keio. Recent research in economics provides evidence that a student's ordinal rank in a peer group affects performance and future career choices in Europe (Elsner et al., 2021; Elsner & Isphording, 2017; Murphy & Weinhardt, 2020), the United States (Denning et al., 2023), China (Yu, 2020), and Japan (Isozumi et al., 2021). This paper also used ordinal rank as the measure of students' performance. In addition to previous findings on the impact of an individual's academic achievement on subsequent donation (Cunningham & Cochi-Ficano, 2002), we newly find that higher ranked students are more likely to donate. However, the rank was not correlated with the amount of donation using the sub-sample of those who donated. Apart from strengthening the community mechanism, the open-academic record system possibly increased the effect of ordinal rank. Concerning the second-stage estimation regarding the amount of donation, the sample size is only 66, which may result in statistical insignificance. Therefore, more rigorous estimation could be conducted if larger sizes of data are used. In the future, we should investigate the effect of the ordinal rank during school days on various types of Keio donations.

# 8. CONCLUDING REMARKS

The open-academic record system has been adopted after 13 years since the establishment of Keio Gijuku in 1858. The system reduced the information asymmetry and strengthened interpersonal relationship. Based on the individual-level dataset constructed connecting three different sets of data, this study examines whether the system is effective in terms of the successful fundraising campaign in 1880 to pull through the difficult situation. Comparing graduates before and after the system was introduced, the main findings are presented as follows: First, relative to the alumni before adopting the system, those who share the academic record under the system are more likely to contribute and their amount of donation tends to be larger. Second, the average class size and their ordinal academic rank in the class are negatively correlated with the likelihood of contribution and with its amount. Third, the higher the academic rank, the higher the likelihood of contribution.

The introduction of the system strengthened the alumni network and role of social image shared by the Keio clan. This typified why the first school fundraising in Japan was successful. Furthermore, the Mita alumni network based on the social image persisted until the present day and displayed the overwhelming ability to raise money to contribute to Keio Gijuku.

The effect of ordinal rank was observed for the probability of contributing but not for the amount of donation. This might be partly due to small sample of those who donated. We should use a larger sample including not only the first donation but also various types of donations collected for Keio to scrutinise whether the ordinal rank fostered the motivation to donate, a case considered as a kind of prosocial behaviour. This is the reimaging issue to be addressed in future studies.

ACKNOWLEDGEMENTS

We thank the anonymous referees and participants of Historical Local Leader Seminar in Keio Gijuku University for many valuable comments. I would like to express my gratitude especially to Prof. Naoko Nishizawa at Keio-Gijuku Fukuzawa-Kenkyu Center for permitting me to use the datasets constructed by members of the centre based on historical records.

CONFLICT OF INTEREST

There is no conflict of interest to be declared in this study.

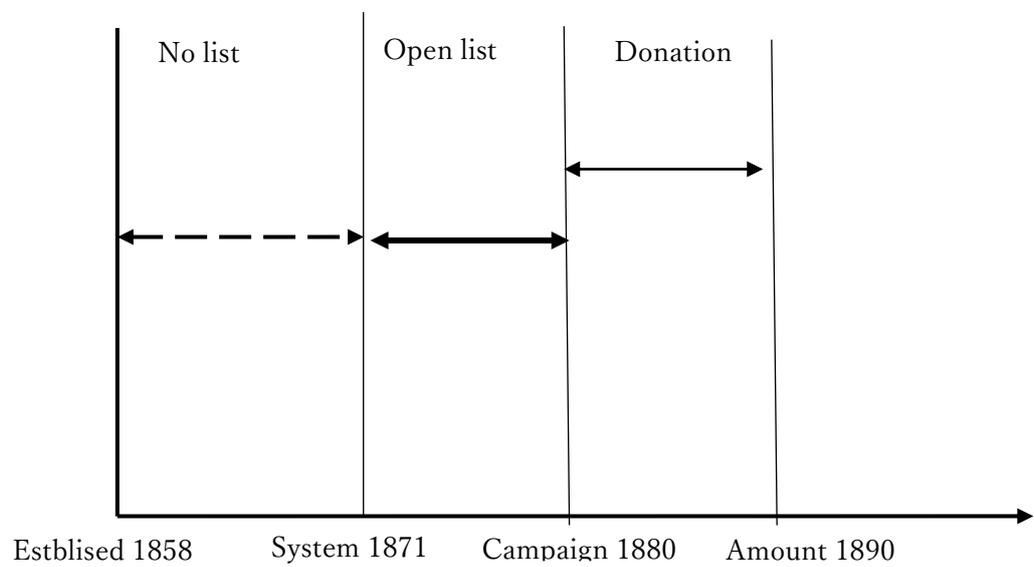

Fig. 1. Time line from 'Keio established', 'the Academic Report System' is introduced, and 'the Raise the contribution to maintenance fund'.

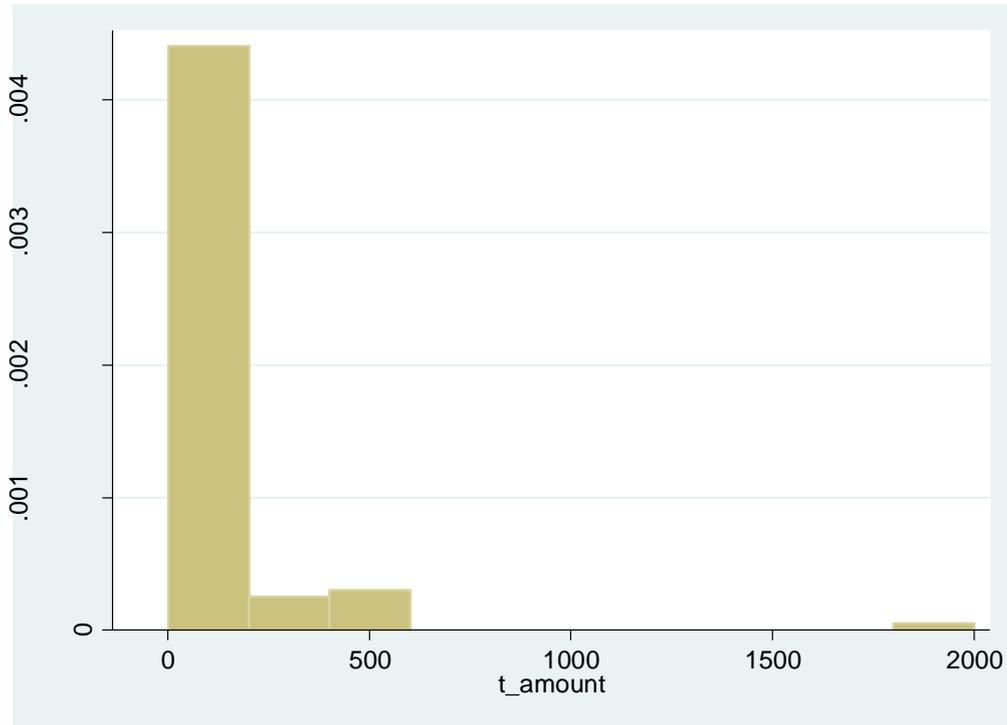

(a) Raw data

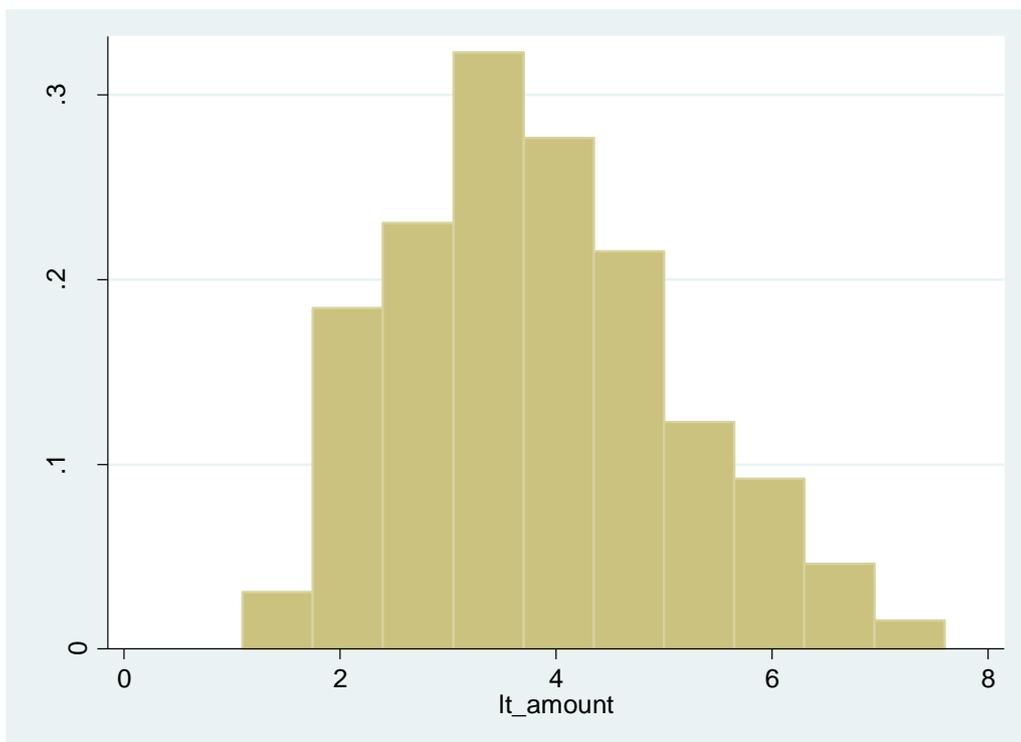

(b) Log form

Fig. 2. Amount of donation

Table 1. Composition of data.
(A) Sample of all individuals who were registered in the list of enrolled students.

| Registered in the list of enrolled students: Obs.=3,053 | | | | |
|---|---|---|---|---|
| The open-academic record system existed: Obs.=2,333 | | | Others | |
| Registered in the report academic record list: Obs.=1,786 | | | Others | |
| Others | Age: Obs.=1,943 | | | Others |
| Others | Entrance Year: Obs.=2,223 | | | Others |

(B) Sample of individuals who donated: Selected from Sample (A).

| Registered in the list of enrolled students: Obs.=100 | | | | |
|---|---|---|---|---|
| The open-academic record system: Obs.=68 | | | Others | |
| Registered in the academic record list: Obs.=66 | | | Others | |
| Others | Age: Obs.=59 | | | Others |
| Others | Entrance Year: Obs.=71 | | | Others |

Note: For both Samples (A) and (B), area sizes are not consistent with number of observations.

Table 2. Description of variables and its basic statistics.
(A) Sample of all individuals who were registered in the list of enrolled students.

| | Description | Mean | S.D. | Obs. |
|---|---|---|---|---|
| *Donation Dummy* | It takes 1 if individuals donated, otherwise 0. | 0.03 | 0.003 | 3503 |
| *Academic Ranking* | Mean values standardised individual's ranking within the class during academic terms. Calculating 'individual's ranking / number of classmates' in each academic term. Then, calculating its mean values for academic terms. | 0.54 | 0.22 | 1786 |
| *Class size* | Number of students in class where respondents belonged | 25.1 | 11.2 | 1786 |
| *Period Length* | Total number of academic terms when individuals learned at Keio | 4.77 | 4.32 | 1786 |
| *Academic Record* | It takes 1 if individual was registered in the academic record list, otherwise 0. | 0.51 | 0.49 | 3503 |
| *Nakatsu* | It takes 1 if individual's hometown was Nakatsu (founder's hometown), otherwise 0. | 0.009 | 0.93 | 3503 |
| *Samurai Class* | It takes 1 if individual belonged to the Samurai Class in the Edo period, otherwise 0. | 0.43 | 0.49 | 3503 |
| *Age* | Individual's age at the starting a drive to raise funds. | 23.2 | 5.61 | 1943 |
| *Entrance Year* | The dominical years when individuals entered the Keio. | 1,873.3 | 4.04 | 2223 |
| *N.A. Age* | It takes 1 if *Age* is not available, otherwise 0. | 0.44 | 0.49 | 3503 |
| *N.A. Entrance Year* | It takes 1 if *Entrance Year* is not available, otherwise 0. | 0.37 | 0.48 | 3503 |
| *Unregistered* | It takes 1 if individual was unregistered in the report card list although the report card system existed when he learned at Keio, otherwise 0. | 0.10 | 0.31 | 3503 |

Notes: Each value is standardised for comparison. * suggests statistical significance at the 10% level.

(B) Sample of individuals who donated.

| | Description | Mean | S.D. | Obs. |
|---|---|---|---|---|
| *Amount* | Amount of donation (yen) | 111.3 | 232.3 | 100 |
| *Academic Ranking* | Mean values standardised individual's ranking within the class during academic terms. Calculating 'individual's ranking / number of classmates' in each academic term. Then, calculating its mean values for academic terms. | 0.42 | 0.19 | 62 |
| *Class size* | Number of students in class where respondents belonged | 22.19 | 7.53 | 62 |
| *Period Length* | Total number of academic terms when individuals learned at Keio | 8.74 | 6.11 | 62 |
| *Academic Record* | It takes 1 if individual was registered in the academic record list, otherwise 0. | 0.62 | 0.48 | 100 |
| *Nakatsu* | It takes 1 if individual's hometown was Nakatsu (founder's hometown), otherwise 0. | 0.09 | 0.29 | 100 |
| *Zaibatsu Work* | It takes 1 if individual worked in the Zaibatsu firm the starting a drive to raise funds, otherwise 0. | 0.14 | 0.34 | 100 |
| *Education Work* | It takes 1 if individual worked as an educator at the starting of a drive to raise funds, otherwise 0. | 0.22 | 0.42 | 100 |
| *Samurai Class* | It takes 1 if individual belonged to the Samurai Class in the Edo period, otherwise 0. | 0.38 | 0.48 | 100 |
| *Age* | Individual's age at the starting a drive to raise funds. | 26.13 | 5.71 | 59 |
| *Entrance Year* | The dominical years when individuals entered the Keio. | 1,870.9 | 4.36 | 71 |
| *N.A. Age* | It takes 1 if *Age* is not available, otherwise 0. | 0.41 | 0.49 | 100 |
| *N.A. Entrance Year* | It takes 1 if *Entrance Year* is not available, otherwise 0. | 0.29 | 0.46 | 100 |

**Table 3.** Mean difference test of *Donation Dummy* in Sample (A)

| (1) | (2) | (1) − (2) |
|---|---|---|
| *Academic Record =1* | *Academic Record =0* | |
| 0.035 | 0.022 | 0.013** |
| *Nakatsu=1* | *Nakatsu=0* | |
| 0.29 | 0.026 | 0.264*** |
| *Samurai=1* | *Samurai=0* | |
| 0.025 | 0.031 | −0.006 |

Note: *, **, and *** indicate statistical significance at the 10%, 5%, and 1% levels, respectively. Information about work categories is only available for Sample (B) and so test cannot be doe using Sample (A).

**Table 4.** Mean difference test of ln (amount of donation) in Sample (B)

| (1) | (2) | (1) − (2) |
|---|---|---|
| *Academic Record =1* | *Academic Record =0* | |
| 3.789 | 3.960 | −0.171 |
| *Nakatsu=1* | *Nakatsu=0* | |
| 4.905 | 3.750 | 1.154*** |
| *Samurai=1* | *Samurai=0* | |
| 3.489 | 4.078 | −0.588** |
| *Zaibatsu Work =1* | *Zaibatsu Work =0* | |
| 4.656 | 3.712 | 0.943*** |

Note: As illustrated in Fig.2, a few outliers existed in amount of donation. Amount of donation is transformed to log value to reduce large variation. *, **, and *** indicate statistical significance at the 10%, 5%, and 1% levels, respectively.

**Table 5.** Mean difference test of academic related variables in Sample (A)

|  | (1) | (2) | (1) − (2) |
|---|---|---|---|
|  | *Donation Dummy =1* | *Donation Dummy =0* |  |
| *Academic Ranking* | 0.415 | 0.549 | −0.134** |
| *Log (Class size)* | 3.056 | 3.177 | −0.121** |
| *Period Length* | 8.742 | 4.635 | 4.106*** |

Note: Class size is transformed to log value to reduce large variation. *, **, and *** indicate statistical significance at the 10%, 5%, and 1% levels, respectively.

**Table 6.** Examination on whether one donated or not (Probit model): Sample (A) is used.

|  | (1) | (2) | (3) |
|---|---|---|---|
| *Academic Record* | 0.095*** (0.041) | 0.127*** (0.048) | 0.130*** (0.050) |
| *Period Length* | 0.002*** (0.0005) | 0.003*** (0.0005) | 0.003*** (0.0005) |
| *Nakatsu* | 0.128*** (0.059) | 0.208*** (0.077) | 0.212*** (0.077) |
| *Ln(Class Size)* | −0.013*** (0.005) | −0.021*** (0.007) | −0.022*** (0.007) |
| *Academic Ranking* | −0.049*** (0.011) | −0.069*** (0.015) | −0.070*** (0.015) |
| *Samurai Class* | 0.002 (0.003) | −0.004 (0.004) |  |
| *Age* | 0.0005 (0.0003) |  |  |
| *Entrance Year* | −0.002*** (0.0005) |  |  |
| *N.A. Age* | 0.011 (0.013) |  |  |
| *N.A. Entrance Year* | −1.00 (0.00) |  |  |
| *Unregistered* | −0.009 (0.005) |  |  |
| Pseudo $R^2$ | 0.149 | 0.107 | 0.107 |
| Observations | 3,053 | 3,053 | 3,053 |

Note: *, **, and *** indicate statistical significance at the 10%, 5%, and 1% levels, respectively. The numbers in parentheses are robust standard errors. The numbers without parentheses are marginal effects.

**Table 7.** Examination on amount of donation in the second stage of the Heckman model:
The first stage is full model shown in Column (1) Table 6.
Dependent variable: Ln(*Amount*)

|  | (1) | (2) | (3) |
|---|---|---|---|
| *Academic Record* | 3.001*** | 2.181* | 2.300** |
|  | (0.947) | (1.122) | (1.146) |
| *Period Length* | 0.018 | −0.014 | −0.013 |
|  | (0.025) | (0.022) | (0.022) |
| *Nakatsu* | 0.799** | 0.226 | 0.188 |
|  | (0.397) | (0.427) | (0.441) |
| *Ln(Class Size)* | −0.896*** | −0.700** | −0.783** |
|  | (0.247) | (0.289) | (0.289) |
| *Academic Ranking* | −0.559 | 0.032 | −0.059 |
|  | (0.74) | (0.839) | (0.857) |
| *Zaibatsu Work* | 0.759** | 0.907*** | 0.859** |
|  | (0.357) | (0.334) | (0.340) |
| *Education Work* | 0.294 | 0.316 | 0.333 |
|  | (0.256) | (0.251) | (0.249) |
| *Samurai Class* | −0.240 | −0.351 |  |
|  | (0.292) | (0.285) |  |
| *Age* | 0.034 |  |  |
|  | (0.029) |  |  |
| *Entrance Year* | −0.044 |  |  |
|  | (0.038) |  |  |
| *N.A. Age* | 0.544 |  |  |
|  | (0.919) |  |  |
| *N.A. Entrance Year* | −82.4 |  |  |
|  | (72.4) |  |  |
| Pseudo R$^2$ | 0.149 | 0.149 | 0.149 |
| Observation | 100 | 100 | 100 |

Note: *, **, and *** indicate statistical significance at the 10%, 5%, and 1% levels, respectively. The numbers in parentheses are robust standard errors. The numbers without parentheses are marginal effects. Samples (A) and (B) are used in the first and the second stages, respectively. In the bottom of this table shows 'Observation' only in the second stage. However, in the first stage, as reported in Table 6, observations are 3,503.